\documentclass[useAMS,usenatbib,usegraphicx]{mn2e}

\newcommand{\e}[1]{ \ensuremath{\times 10^{#1}}}
\newcommand{\psrbs}{PSR~B1259-63/SS~2883}
\newcommand{\psrb}{PSR~B1259-63}

\title[The effect of the infrared excess on IC gamma-ray production]{The effect of the infrared excess from the Be star on inverse Compton gamma-ray production in PSR~B1259-63/SS~2883}

\author[B. van Soelen and P.~J. Meintjes]
  {B.~van~Soelen\thanks{E-mail:
vansoelenb@ufs.ac.za},
  P.~J.~Meintjes\\Department of Physics, University of the Free State, South Africa, 9300\\}
\date{Released 2010 Xxxxx XX}

\begin{document}

\maketitle

\begin{abstract}

The binary system \psrb\ consists of a 48~ms pulsar in a 3.4 year orbit around a Be star and unpulsed TeV gamma-ray emission has been detected near previous periastron passages. A likely source of the gamma-rays is the inverse Compton up-scattering of target photons from the Be star by the ultra-relativistic electron/positron pulsar wind in the region of the plerion shock front.  In this study the effect of the infrared emission from the Be star's circumstellar disc on inverse Compton gamma-ray production in \psrb\ is investigated by considering an isotropic photon/electron approximation. The modified photon density distribution is determined by using a curve of growth method fitted to previous optical and infrared observations.  The inverse Compton scattering rate is calculated using the modified photon distribution and the exact scattering equation. It is shown that including the infrared emission increase the GeV  gamma-ray flux by a factor $\ga 2$. 

\end{abstract}


\section{Introduction}

Be stars -- defined as B-type stars showing or having showed emission lines instead of absorption lines in the Balmer spectrum \citep{collins87} -- are known to produce an infrared (IR) flux which is higher than predicted by Kurucz stellar atmosphere models \citep{kurucz79}.   It has long been speculated that Be stars possess an extended circumstellar envelope or disc \citep[e.g.][]{struve31} and optical interferometry observations confirm the presence of an extended circumstellar envelope which is symmetrical around the rotation axis \citep[see e.g.][]{quirrenbach94}.
These circumstellar discs act as reverse accretion discs as they are regions (confined to a disc structure) of increased stellar outflow.  The free-free and free-bound emission which occurs in the disc is presumably responsible for the observed IR excess.  These discs also show great variability and grow and shrink over periods of hundreds to thousands of days.  For a review of classical Be stars, see e.g.\ \citet{porter03}. 

An important Be X-ray Pulsar binary (Be-XPB) example is the TeV gamma-ray system \psrb\ which consists of a 48~ms pulsar in  orbit around a Be star, SS~2883, and was initially detected as part of a radio survey of the southern Galactic plane \citep{johnston92,johnston94}.  The orbit is eccentric ($ e \approx 0.87 $) and unpulsed TeV gamma-ray emission has been detected by HESS\footnote{The High Energy Stereoscopic System, located near the Gamsberg Mountain in Namibia, consists of four Cherenkov imaging telescopes.} close to the periastron passage \citep{aharonian05,aharonian09}.  The circumstellar disc is mis-aligned to the orbital plane and the pulsar is believed to pass  through it twice: before and after periastron.  
Previous observations have also detected unpulsed radio emission near periastron \citep[e.g][]{johnston99} and the system has been observed at X-ray frequencies across the whole orbit \cite[see e.g.][]{cominsky94full,kaspi95,chernyakova06,chernyakova09}.
These observations show that the unpulsed emission is variable at all wavelengths with respect to the orbital period.

The system is powered by the spin-down luminosity of the pulsar and the unpulsed radiation is believed to originate from a stand-off shock front between the pulsar and stellar wind; extensive modelling has been presented in \citet{tavani97}.  In this model the X-rays are produced via synchrotron radiation of ultra-relativistic electrons (Lorentz factor $\gamma \sim 10^6$) and the gamma-ray emission through the inverse Compton (IC) scattering of target photons from the Be star.  The ratio between the X-ray and gamma-ray flux is dependent on the radiative and adiabatic cooling times of the post-shocked wind and IC cooling was predicted to increase near periastron.  IC cooling of the pre-shocked wind was also suggested by \citet{ball00}.

A slower wind model has also been proposed \citep{chernyakova99,chernyakova00} where the pulsar wind has a Lorentz factor of $\gamma \sim 10-100$ and up-scatters the target photons from the Be star via IC processes to X-ray energies.  The TeV gamma-rays are then explained via the interaction of the pulsar with the circumstellar disc and are the result of proton-proton collisions, IC scattering and bremsstrahlung emission \citep{kawachi04,chernyakova06}.  

If the TeV gamma-rays are the result of IC scattering of target photons from the Be star, this implies that electron energies are of the order of TeV and the observed TeV gamma-rays are the result of scattering which occurred in the Klein--Nishina limit.  While previous models have considered IC scattering in the general case, there are currently only TeV gamma-ray observations of \psrb\ and the emphasis has been to model the TeV emission which will occur in the Klein--Nishina limit.  Previous models have not considered how the scattering of the IR excess from the circumstellar disc, occurring in the Thomson limit, will influence the gamma-ray production.

It can be shown \cite[e.g.][]{bg70} that IC scattering between a relativistic electron with a Lorentz factor $\gamma$ and a photon with a frequency $\nu$ will occur in the classical Thomson limit if 
\begin{equation}
\nu  \ll \frac{m_\rmn{e} c^{2}}{h \gamma}, 
\label{eqn:thomlimit}
\end{equation}
where $h$, $m_\rmn{e}$ and $c$ represent Planck's constant, the electron mass and the speed of light respectively.  To explain the results of previous HESS observations of \psrb, \citet{aharonian05} proposed that electrons are accelerated in the shock between the pulsar ($\gamma \sim 10^{6} - 10^{7}$) and stellar wind, resulting in the IC scattering of stellar photons.  This scattering will occur in the Thomson limit for target photons with frequency $\nu \ll 10^{14} (\gamma/10^{6})^{-1}$~Hz, i.e. radio to IR.
The IR excess provides an additional reservoir of soft target photons for IC scattering and an attractive framework to explain gamma-ray production through IC scattering in the Thomson limit, where the cross-section is significantly higher than the Klein--Nishina limit.  A more detailed discussion, especially focussing on the IC spectral properties,  will be presented in section~\ref{ic}.

Since the IR excess is tied to the growth and decay of the circumstellar disc, the size of the disc near periastron will influence gamma-ray production not only through the constraints on the plerion shock, but also through IC scattering in the Thomson limit.

This study will discuss the effect of the IR excess on IC gamma-ray production in Be-XPBs using \psrb\ as a trial case. Results of modelling the full IC scattering spectrum (Thomson and Klein--Nishina limit) of an isotropic distribution of photons and electrons will be presented here as an initial argument towards the importance of the contribution of the IR excess.
The additional effects of the different electron cooling processes in the plerion shock front and the changing scattering angle, due to orbital motion, will not be discussed at this time as the goal of this paper is to validate the importance of the IR excess on IC gamma-ray production. In order to include the IR excess the target photon distribution will be determined using the curve of growth method proposed by \citet{lamers84} (referred to hereafter as the COG method) fitted to previous optical and IR observations.  

An outline of previous observations and studies of \psrb\ will be presented (section~\ref{psrb}) before discussing the COG method (section~\ref{be-stars}) and IC scattering (section~\ref{ic}). Section~\ref{sec:modelling} will outline the modelling undertaken in this study and  sections~\ref{sec:discussion} and \ref{sec:conclusion} will discuss the results and implications thereof.

\section{PSR B1259-63/SS 2883}
\label{psrb}

While this paper focuses only on one aspect of the the multiwavelength emission from \psrb, a complete model needs to take into account all wavelengths to accurately explain the radiative processes in the system. Below a brief summary is given of previous observations and modelling of the system which is relevant to the discussion presented in this paper.

There have been few photometric observations of SS~2883, the optical companion to \psrb, but some measurements are available in bright star catalogues (Table~\ref{tab:tab1}).  For example, \citet{westerlund89} presented photometric and visual extinction measurements for SS~2883 as part of their catalogue of OB stars near the Southern Coalsack.  Comparison of the catalogue's observations show little change in the optical magnitude.
From this and other observations \citet{johnston94} confined the spectral range of SS~2883 to a O9 - B2 type star and assumed a spectral type B2e, with $M_* \simeq 10$~M$_{\odot}$ and $R_* \simeq 6$~R$_\odot$. \citet{johnston94} also undertook spectroscopic observations of the system and observed double peak H$\beta$, H$\gamma$, H$\delta$ and He~I~5876 emission lines.  Analysis showed that the H$\beta$ emission occurred at a radius of $8.5~R_*$, assuming a Keplerian circumstellar disc, which infers a circumstellar disc radius $R_\rmn{disc}> 8.5~R_*$.  

There are also few observations in the mid--to--near IR, but observations are available from the \textit{2MASS} \citep{2mass} and \textit{MSX}\footnote{Data from \textit{2MASS} and the \textit{Midcourse Space Experiment} (\textit{MSX}) are available on-line through the NASA/IPAC Infrared Science Archive.} \citep{price01} missions.  The IR data is shown in Table~\ref{tab:tab2}.  Due to the large uncertainty in the C-band measurement it was not considered when modelling this system.

In addition to the pulsed radio signal, an unpulsed radio signal has been detected from the system close to periastron \cite[see e.g.][]{johnston05}.  Observations of the system during consecutive periastron passages show an eclipse of the pulsar -- and a decrease in the unpulsed synchrotron emission -- roughly 20 days before and after periastron passage as it passes through and behind the circumstellar disc.  For example, the pulsar eclipse lasted from 16 days before until 18 days after periastron during the 2000 passage \citep{connors02} and from 15.8 days before until 16.1 days after periastron during the 2004 passage \citep{johnston05}. At $~16 - 20$~days from periastron the binary separation is $\approx 40 - 50~R_*$, assuming $M_\rmn{pulsar} = 1.4~\rmn{M_\odot}$, which implies that the circumstellar disc must at least extend to these distances. 
While the synchrotron radio observations show many similarities there are marked differences in the flux levels during consecutive periastron passages, attributed to variations in the circumstellar disc and material \cite[see e.g. fig.~3 and discussion in][]{johnston05}. 

\begin{table*}
\begin{minipage}{160mm}
\caption{Optical magnitudes and colours for SS~2883}
\label{tab:tab1}
\begin{tabular}{llllllllll} \hline
Catalogue & LSS & Spectral Type & V & B-V & U-B & (U-B)$_0$ & A$_\nu$ & M$_\nu$(U-B)$_0$ \\ \hline
\citet{westerlund89} & 2883 & OB+ce,1e,h & 10.01 & 0.754 & -0.506 & -1.261 & 3.25 & -4.6 \\
\citet{klare77} & 2883 & & 10.04 & 0.74 & -0.44 & & & \\ 
\citet{schild83}  & 2883 & & 10.05 & 0.72 & -0.47 & & & \\ 
\citet{drilling91}   & 2883 & & 10.07 & 0.73 & -0.47 & & & \\ \hline
\end{tabular}
\end{minipage}

\end{table*}

\begin{table}
\caption{IR data for \psrb\ from \textit{2MASS} (J,H,Ks) and \textit{MSX} (A,C). }
\label{tab:tab2}
 \begin{tabular}{cccc} \hline
Band & Wavelength & Magnitude & Flux \\
&	$\mu$m & & Jy \\ \hline
J & 1.235 & $8.026\pm0.027$ & \\
H & 1.662 & $7.699 \pm 0.055$ & \\
Ks & 2.159 & $7.248 \pm 0.020$ & \\
A & 8.28& &$0.2675 \pm 0.0136$ \\
C & 12.13 & & $1.087 \pm 0.5859$\\ \hline
\end{tabular}
\end{table}

As a result of the alignment of the unpulsed radio, X-ray, and the 2004 HESS light curves into two peaks around the same two orbital phases (before and after periastron passage) \citet{chernyakova06} suggested that the peaks in radio and X-ray, and the production of gamma-rays, was due to the pulsar passing through the disc.  This implied a circumstellar disc with a half-opening angle $18\fdg5$, inclined $70\degr$ to the orbital plane.  This suggested disc geometry is, however, inconsistent with the radio eclipse \citep{khangulyan07}.  In addition, the detection of TeV gamma-rays 47 days before the 2007 periastron passage, mis-aligned to the orbital phase position proposed above, suggests that IC scattering still plays a dominant role in gamma-ray production \citep{aharonian09}. 

The exact position of the circumstellar disc and details behind the observed unpulsed radio, X-ray and gamma-ray emission (slow wind, $\gamma \sim 10^2$, or fast wind, $\gamma \sim 10^6$) remains unsolved and the observations of \psrb\ by \textit{Fermi} around the next periastron passage may help answer these questions.  
In this study we assume that the electrons have a Lorentz factor $\gamma \sim 10^6$ and that the X-ray emission is the result of synchrotron radiation while the gamma-rays are produced through IC scattering.  Under this assumption we present results of the full IC scattering of the Be star's photon spectrum taking into account the IR excess produced by the disc.

\section{The Lamers \& Waters Curve of Growth Method} 
\label{be-stars}

In this paper the COG method proposed by \citet{lamers84} has been chosen to model the IR excess from the circumstellar disc. This method allows for some prediction of the variability of the IR excess that will occur because of the changing size of the disc. The method, as discussed in \citet{lamers84}, \citet{waters86} and \citet{telting98}, is summarised below.

Under the COG method it is assumed that the circumstellar disc has a half-opening angle $\theta$, extends to a radius $R_{\rmn{disc}}$,  and has a power-law density profile that decreases with distance as 
\[
 \rho(r) = \rho_0 \left(\frac{r}{R_*}\right)^{-n}.
\]
For a Be star-disc system viewed face-on the flux ratio between the disc and star is given by \citep{telting98}\footnote{In order to keep the notation consistent within this paper, $R_\rmn{disc}$ is in units of centimetres and the integration in equation~(\ref{equ:irratio}) is between $0 -(R_\rmn{disc}/R_*$). This is different to equation (6) in \citet{telting98} where the integration is between $0 - R_\rmn{disc}$, but $R_\rmn{disc}$ is in units of stellar radii.  In addition, the optical depth, $\tau(q)$, in \citet{telting98} (their equations (2) \& (6)) is equivalent to $2\tau_\nu(q)$ as is given here, when the disc is viewed face on.}
\begin{equation}
 \frac{F_{\nu,\rmn{disc}}}{F_{\nu,*}} = \frac{B_\nu (T_\rmn{disc})}{I_{\nu,*}} \int_1^{R_\rmn{disc}/R_*} \left[1-e^{-2\tau_\nu(q)}\right] \, 2q \, dq,
\label{equ:irratio}
\end{equation}
where $F_{\nu,*}$ is the flux from the star and $F_{\nu,\rmn{disc}}$ is the flux from the disc, $B_\nu(T_\rmn{disc})$ is the Planck blackbody function, $T_\rmn{disc}$ is the temperature of the disc, $I_{\nu,*}$ is the flux from the appropriate Kurucz model for the star, and $q$ is the impact parameter in units of the stellar radius. The optical depth, $ \tau_\nu(q)$, is given by \citep{waters86}
\[
 \tau_\nu(q) = E_{\nu, \rmn{disc}} \, q^{-2n+1} C(n,\theta),
\]
where
\[
 C(n,\theta) = \int_0^\theta \cos^{2n-2} y \, dy,
\]
where $\theta$, the half-opening angle, is in radians and $E_{\nu, \rmn{disc}}$ is defined as the optical depth parameter for the disc.  This parameter is given by
\[
 E_{\nu,\rmn{disc}} = X_\lambda X_{*d}
\]
 where
\begin{eqnarray*}
 \lefteqn{ X_\lambda= \lambda^2 \{ ( 1 - e^{-h\nu/kT_\rmn{disc}} )/(h\nu/kT_\rmn{disc})\}} \\ 
& &\times \{g(\nu,T_\rmn{disc}) + b(\nu,T_\rmn{disc})\}
\end{eqnarray*}
and
\[
 X_{*d} = 4.923 \times 10^{35} \overline{z^2} T_\rmn{disc}^{-3/2} \mu^{-2} \xi \rho_0^2 \left(\frac{R_*}{R_\odot}\right),
\]
where $\lambda$ is the wavelength, $\nu$ is the frequency, $k$ is the Boltzmann constant,  $g(\nu,T_\rmn{disc}) + b(\nu,T_\rmn{disc})$ is the sum of the free-free and free-bound gaunt factors,\footnote{The free-free + free-bound gaunt factors published in tabular form in \citet{waters84} for wavelengths between $10^{-4} - 6$~cm, were used in this study.} $\overline{z^2}$ is the mean of the squared atomic charge, $\mu$ the mean atomic weight (units of proton mass) and $\xi$ the ratio of the number of electrons to the number of ions.  

The COG method can then be used to model a Be star's optical/IR flux by modifying a Kurucz model atmosphere at IR wavelengths with the flux ratio given by equation~(\ref{equ:irratio}), which is calculated by fitting the five parameters $n$, $X_*$, $R_\rmn{disc}$, $T_\rmn{disc}$ and $\theta$ to IR data. The Kurucz model atmosphere is chosen by a least-squares fit to the optical data, which is assumed to be less affected by the IR excess.  In practise, however, assumed values are taken for the disc temperature (taken to be a fraction of the stellar temperature) and half-opening angle. The temperature ratio, $T_\rmn{disc} = 0.8~T_*$, used in \citet{lamers84}, \citet{waters86} and \citet{telting98}  has been argued to be too high by \citet{millar99} from their analysis of the energy gain/loss rate in four Be stars.  For this reason a lower temperature ratio of $T_\rmn{disc} = 0.5~T_*$ will be used in this study instead.  From the pulsar eclipse a last constraint can be placed on the disc radius and $R_\rmn{disc} = 50~R_*$ will be used in most instances.  This corresponds to the separation between the pulsar and Be star at the start of the pulsar eclipse. In fact, from polarisation and rotation measurements made around the 1994 periastron passage, \citet{johnston96} suggested that the pulsar begins to pass into the disc $\sim 100$~days before and after periastron at a distance of $\sim 150 R_*$.

With this additional constraint on the disc radius there are only two free parameters ($n$ \& $X_*$) which must be fitted to match the IR excess.

\section{Inverse Compton Scattering}
\label{ic}

A very brief discussion of IC scattering,  presenting only the results relevant for this study, will now be given. For detailed reviews see, e.g.\ \citet{bg70} and \citet{rybicki04}.

IC scattering involves the up-scatter of photons to higher energies by relativistic electrons. It can be shown \cite[e.g.][]{bg70} that
the interaction between relativistic electrons with velocity $\beta = v/c$ and photons with energy $h\nu$ results in the photons being up-scattered to energies 
\[
\epsilon_{\gamma} \approx 4 \gamma^{2} h \nu
\]
in the Thomson limit ($\gamma h \nu <<  m_{\rmn e} c^{2}$), and
\[
\epsilon_{\gamma} \approx \gamma m_{\rmn e} c^{2} 
\]
in the extreme Klein--Nishina limit ($\gamma h \nu >> m_{\rm e} c^{2}$ and $\beta \rightarrow 1$). Although in the Thomson limit the characteristic energy of the scattered photon is increased by a factor $\sim \gamma^2$, it is still small compared to the energy of the relativistic electron. In the case of scattering in the ultra-relativistic limit the scattered photon energies are limited by the electron energies.

In a single scattering event (a single electron scattering a single photon) the energy of the electron before scattering is given by $\gamma m_e c^2$ and the energy of the photon, before and after scattering, by $\epsilon$ and $\epsilon_1$ respectively.   
By expressing the energy of the scattered photon in units of the electron rest mass energy, i.e.  $E_1 = \epsilon_1/\gamma m_e c^2$, the scattering rate per unit energy, for a single electron interacting with an isotropic photon density distribution $n(\epsilon)$ (i.e. number density of photons per unit energy), is expressed as \citep{bg70}
\begin{eqnarray}
\nonumber \lefteqn{\frac{dN_{\gamma,\epsilon}}{dt\,dE_1} = \frac{2 \pi r_0^2 m_e c^3}{\gamma} \frac{n(\epsilon)\, d\epsilon}{\epsilon} }\\
& &\times \left[ 2q \ln q + ( 1+2q)(1-q) + \frac{1}{2} \frac{\left(\Gamma_\epsilon q\right)^2}{1+\Gamma_\epsilon q} (1-q) \right],
\label{eq_bg_scattering_rate}
\end{eqnarray}
where 
\[
 \Gamma_\epsilon = \frac{4 \epsilon \gamma}{m_e c^2}, \qquad q = \frac{E_1}{\Gamma_\epsilon ( 1 - E_1)},
\] 
and $r_0$ represents the classical electron radius.  This is the exact expression for the scattering rate and is appropriate for all energies, provided that $\gamma \gg 1$, with $E_1$ within the range
\[
 \frac{\epsilon}{\gamma m_e c^2} \leq E_1 \leq \frac{\Gamma_\epsilon}{1+\Gamma_\epsilon}.
\]
In equation~(\ref{eq_bg_scattering_rate}), $n(\epsilon) \, d\epsilon$ is the differential photon number density and in the case of a blackbody distribution of photons is \citep[e.g.][]{bg70}
\[
 n(\epsilon) = \frac{1}{\pi^2 (\hbar c)^3} \frac{\epsilon^2}{e^{\epsilon/kT}-1}.
\]

To extend the calculations to multiple electron scatterings an electron distribution function needs to be considered. This distribution is normally expressed as a power law function e.g.
\begin{eqnarray}
 \nonumber N_e(\gamma) & = & K_e \gamma^{-p}, \quad \gamma_\rmn{min} < \gamma < \gamma_\rmn{max}, \\
& = & 0,    \qquad \qquad \rmn{elsewhere,}
\label{eq:electron_distribution}
\end{eqnarray}
where the distribution is confined to the energy region $\gamma_\rmn{min}$ to $\gamma_\rmn{max}$.
The total scattering rate per energy for an electron distribution is then found by intergrating over the initial electron and photon energies, i.e.
\begin{equation}
 \frac{dN_\rmn{total}}{dt d\epsilon_1}  = \int_\epsilon \int_\gamma N_e(\gamma) \times \left( \frac{1}{\gamma m_ec^2}  \frac{dN_{\gamma,\epsilon}}{dt\,dE_1} \right) \, d\gamma \, d\epsilon.
\label{equ:dN_total}
\end{equation}

\section{Modelling the IC scattering}
\label{sec:modelling}

\subsection{Modelling the Infrared Excess}
\label{sec:modelling_cog}

The optical and infrared data used to model \psrb\ in this paper is taken from \citet{westerlund89}, \textit{2MASS} and \textit{MSX} (Tables~\ref{tab:tab1} \& \ref{tab:tab2}). The IR and optical data were de-reddened with the {\sc dipso} software package using the extinction measurement given in \citet{westerlund89} and following the assumptions of \citet{johnston94}.

The initial Kurucz atmosphere fits were done using the \citet{kurucz92} model atmospheres obtained from the Space Telescope Science Institute\footnote{\emph{http://www.stsci.edu/science/starburst/Kurucz.html}}.  The tables only begin at a minimum frequency of $\sim 3.3\e{13}$~Hz and needed to be extended further into the IR regime.  The last few data points lie within the  Rayleigh-Jeans limit ($h\nu \ll kT$) for a stellar temperature of $\sim 20\,000$~K, and since in this limit $\log I_{\nu,*} \propto \log \nu$, the data points were extended to $\nu \approx 5.2\e{9}$~Hz (or $\lambda \sim 6$~cm) using a straight line fit.

The stellar temperature, $T_*$, and effective gravity, $\log g$, were determined by finding the best least squares fit of the Kurucz model atmospheres to the optical data above $3 \times 10^{14}$~Hz (in order to limit the influence of the IR excess on the optical fit). The decrease in flux due to distance to the source from Earth ($d$) was also accounted for by scaling the intensity with the parameter $Y_\rmn{shift} = \log(\pi R_*^2) - \log(d^2)$ \citep{telting98}. 
A best fit temperature and effective gravity of $T_{*} = 25000$~K and $\log~g = 3.5$, respectively, were found, consistent with an early B type star. 

\begin{figure}
 \includegraphics[scale=.45]{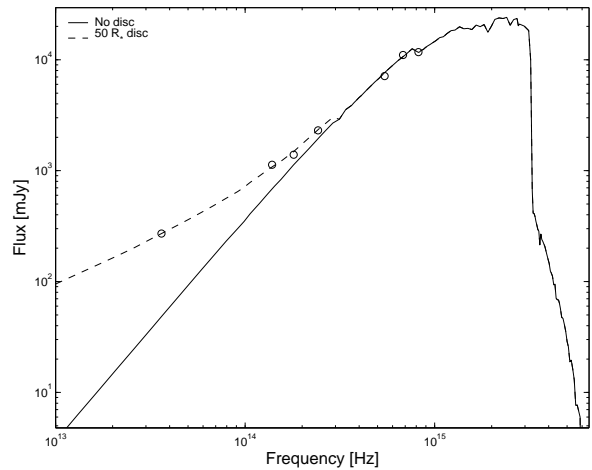}
 % predicted_flux_VLT.eps: 1179666x1310738 pixel, 0dpi, infxinf cm, bb=
 \caption{A Kurucz stellar atmosphere (solid line) and the modification create by the COG method (dashed line) fitted to data for \psrbs.  The open circles are the optical and IR data points.}
 \label{fig:fig1}
\end{figure}

The IR excess was then fitted to the IR data using the COG method. The disc temperature was chosen to be $T_\rmn{disc} = 0.5~T_* = 12500$~K and the disc radius $R_\rmn{disc} = 50~R_{*}$ as was inferred from the pulsar eclipse. 
In this paper the convention used by \citet{telting98} was followed and the half-opening angle was held at $\theta = 5\degr$.   The remaining two parameters ($n$ and $X_*$) were simultaneously fitted with the Levenberg-Marquardt method using the implementation in the {\sc oracle} software package. This produced a best fit of $n=2.37$ and $\log X_* = 7.87$ and the resulting fit is plotted in Fig.~\ref{fig:fig1}. Fits using the larger half-opening angle ($\theta \sim 18\fdg5$) suggested by \citet{chernyakova06} were also considered but produced no noticeable effect on the fitted optical/IR spectrum.  The values are summarised in Table~\ref{tab:tab3}.

As a futher check on these model simulations, COG fits were produced using the data and parameters given in \citet{waters86} and \citet{telting98} for $\delta$~Cen and X~Persei respectively.  These models were compared to the published results and were found to match extremely well.

\begin{table}
 \caption{Optical and IR fit to PSR~B1259-63}
\label{tab:tab3}
% use packages: array
\begin{tabular}{lll} \hline
\textit{Optical fit} & & \\
$T_\rmn{eff}$ & 25000~K &  \\ 
$\log g$ & 3.5 & × \\ 
$Y_\rmn{shift}$ & 19.64 & × \\ 
\\
\textit{Curve of growth fit} & × & × \\ 
$n$ & 2.37 & 2.37 \\ 
$\log X_*$ & 7.87 & 7.32 \\ 
$R_\rmn{disc}$ & $50~R_*$ & $50~R_*$ \\ 
$T_\rmn{disc}$ & 12500~K & 12500~K \\ 
$\theta$ & $5\degr$ & $18\fdg5$ \\ \hline
\end{tabular}
\end{table}

\subsection{Modelling the IC gamma-ray spectrum}
\label{sec:modelling_ic}

The total number of IC scatterings is calculated from equation~(\ref{equ:dN_total}) and the photon distribution $n(\epsilon)$ is derived from the predicted flux as is given by the COG fit.  This is different to previous models where either a blackbody or mono-energetic photon spectrum was assumed.  The integration over $\epsilon$ was performed for the range $\epsilon = 3.45\times 10^{-17} - 4.23\times10^{-10}$~ergs ($\sim 5.2\times10^9 - 6.4\times10^{16}$~Hz). At these limits, the ratio $n(\epsilon_\rmn{limit}) / \rmn{\textbf{max}}(n(\epsilon))$ is $1.13\times 10^{-5} $ and $ 1.60 \times 10^{-39}$ respectively. The integration over $\gamma$ was performed between the given $\gamma_\rmn{min}$ and $\gamma_\rmn{max}$ limits for each electron distribution case.

The electron distribution, $N_e$, in the IC calculation depends on where the electrons are assumed to originate, and what additional cooling processes are applicable.  The examples of pre-shocked, post-shocked adiabatic cooling  and post-shocked radiative cooling are presented below.  Since the aim of this study is to show the relative change in the flux of the gamma-ray spectrum, the models are not calibrated to fit the HESS data and two simplifying assumptions have been made in the analysis below; the value of $K_e=1$ was adopted for all cases and the photon density has not been scaled to the distance of the pulsar.  Scaling the photon density simply decreases the overall gamma-ray flux by a constant value and does not change the relative shape.

\subsubsection{Pre-shocked electrons}

\begin{figure}
 \includegraphics[scale=.45]{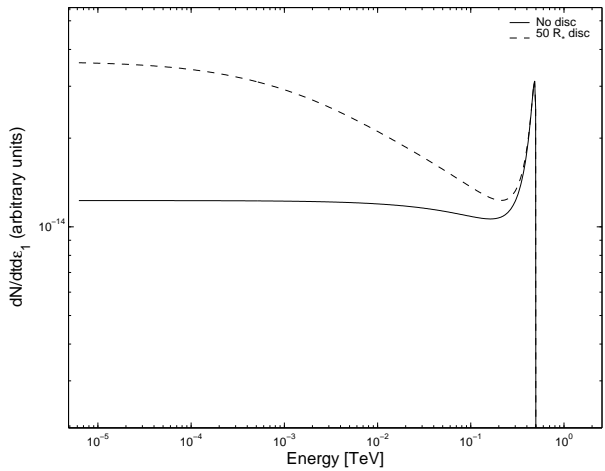}
\caption{The predicted total scattering rate, $dN_{tot}/dt\,d\epsilon_1$, calculated for a mono-energetic electron $\gamma = 10^6$.  The solid line is the expected gamma-ray flux without an IR excess and the dashed line shows the increased gamma-ray flux when the IR excess is included.}
\label{fig:fig2}
\end{figure}

The possibility that IC scattering may occur in the pre-shocked region surrounding the pulsar has been discussed by \citet{ball00}.  In such a scenario the pulsar wind would have an approximate mono-energetic distribution around $\gamma \sim 10^6$.  Fig.~\ref{fig:fig2} presents this scenario, which was calculated from equation~(\ref{equ:dN_total}) without integrating over $\gamma$.  The scattering rate shows a significant increase when the IR excess is included.

\subsubsection{Post-shock electrons - Adiabatic cooling}

\begin{figure}
 \includegraphics[scale=.45]{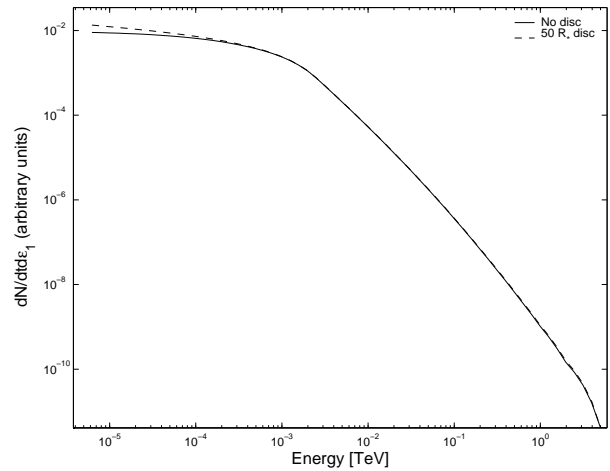}
\caption{The predicted total scattering rate, $dN_{tot}/dt\,d\epsilon_1$, calculated using $p = 2.2$ for the range $\gamma = 10^4 - 10^7$. The electron distribution follow equation~(\ref{eq:electron_distribution}). The solid line is the expected gamma-ray flux without an IR excess and the dashed line shows the increased gamma-ray flux when the IR excess is included.}
\label{fig:fig3}
\end{figure}

\begin{figure}
 \includegraphics[scale=.45]{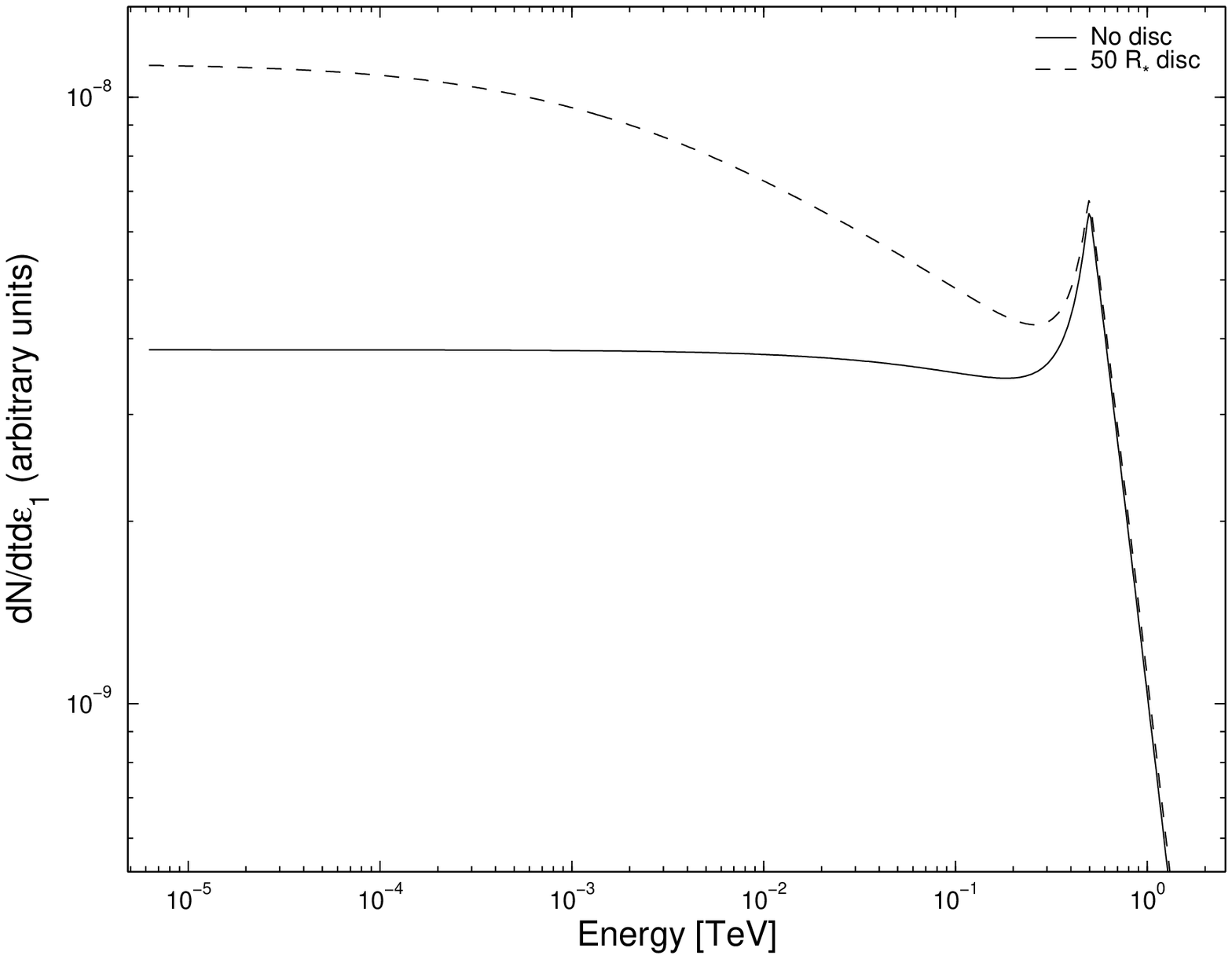}
\caption{The predicted total scattering rate, $dN_{tot}/dt\,d\epsilon_1$, calculated using $p = 2.2$ for the range $\gamma = 10^6 - 10^7$. The electron distribution follow equation~(\ref{eq:electron_distribution}).  The solid line is the expected gamma-ray flux without an IR excess and the dashed line shows the increased gamma-ray flux when the IR excess is included.}
\label{fig:fig4}
\end{figure}

If cooling is dominated by adiabatic cooling in the post-shock region the electron distribution is assumed to be a power law ($N_e  = \gamma^{-p}$).  The HESS observations of \psrb\ suggest that $p \approx 2.2$ \citep{aharonian05} and this was adopted as a first approximation.
Two energy ranges for $\gamma$ were considered, $\gamma = 10^4 - 10^7$ and $\gamma = 10^6 - 10^7$, encompassing a broad and narrow electron distribution.  The resulting scattering rates are shown in Fig.~\ref{fig:fig3} and Fig.~\ref{fig:fig4} respectively.
For the broad electron distribution there is negligible influence on the gamma-ray flux when the IR excess is included.  For the higher energy, narrow band, electron distribution, this is not the case and there is a significant increase in the scattering rate.

\begin{figure}
 \includegraphics[scale=.45]{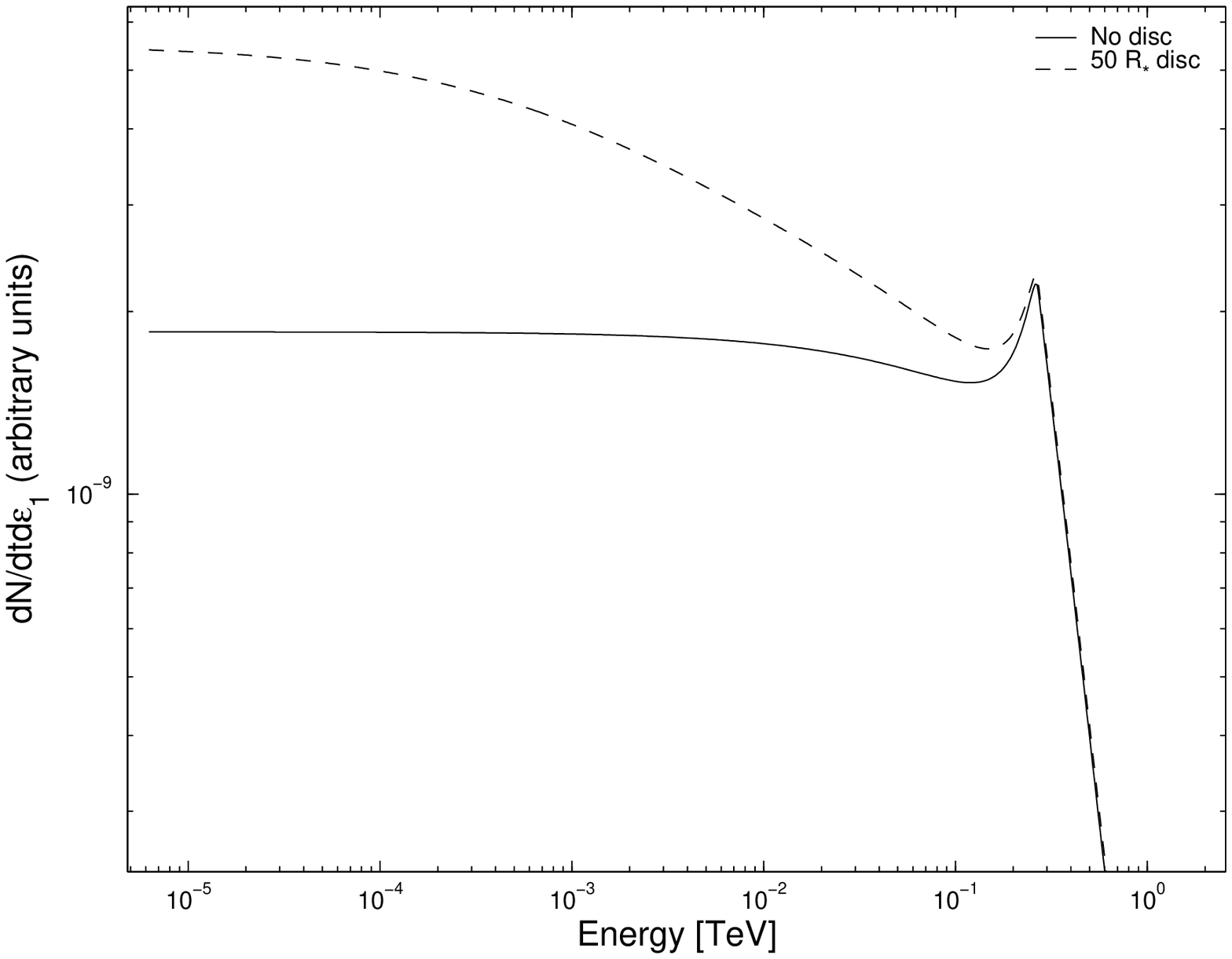}
\caption{The predicted total scattering rate, $dN_{tot}/dt\,d\epsilon_1$, calculated using $p = 2.4$ for the range $\gamma = 5.4\times10^6 - 5.4\times10^7$ \citep[see][]{kirk99}. The electron distribution follow equation~(\ref{eq:electron_distribution}).  The solid line is the expected gamma-ray flux without an IR excess and the dashed line shows the increased gamma-ray flux when the IR excess is included.}
\label{fig:fig5}
\end{figure}

A third example of a predominately adiabatic cooled electron distribution was taken from \citet*{kirk99}.  Here the authors considered $p=2.4$ and $\gamma = 5.4\times10^5 - 5.4\times10^7$ and the spectrum is comparable to the HESS observations (see fig.~7 in \citet{aharonian05}).  The resulting scattering rate (assuming $K_e = 1$) is shown in Fig.~\ref{fig:fig5}.  There is a significant increase in the scattering rate due to the IR excess, though slightly less than is presented in Fig.~\ref{fig:fig4}.  The peak in the gamma-ray spectrum also occurs at a lower energy.

\subsubsection{Post-shock electrons - Radiative Cooling}

\begin{figure}
 \includegraphics[scale=.45]{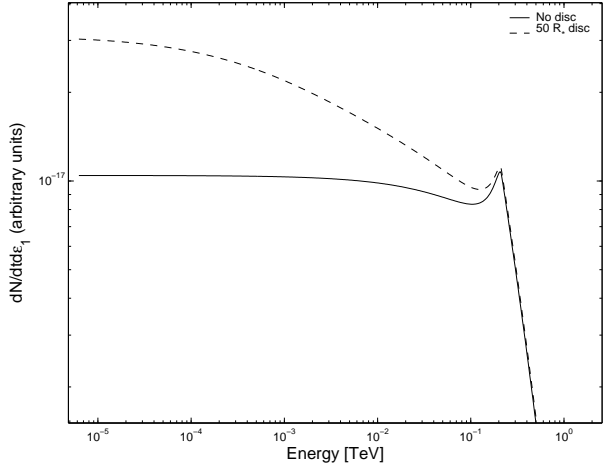}
\caption{The predicted total scattering rate, $dN_{tot}/dt\,d\epsilon_1$, calculated using the electron spectrum from \citet{kirk99} for dominate radiative cooling; $B=0.32$~G and $p = 1.4$ for the range $\gamma = 4.3\times10^5 - 4.3\times10^7$.  The solid line is the expected gamma-ray flux without an IR excess and the dashed line shows the increased gamma-ray flux when the IR excess is included.}
\label{fig:fig6}
\end{figure}

When the electrons in the post-shocked region also cool via synchrotron radiation, the electron distribution can not be considered as a simple power law. While the post-shocked electrons initially have a power law distribution this is modified as the electrons lose energy via synchrotron radiation.  An example of dominant radiation cooling, with a magnetic field strength of $B=0.32$~G, was considered, also taken from \citet{kirk99}. The initial electron spectrum was assumed to have an index of $p = 1.4$ and the energy range was $\gamma = 4.3\times10^5 - 4.3\times10^7$. This modified electron distribution was used to model the scattering rate shown in Fig.~\ref{fig:fig6}.  The details of the modification are beyond the scope of this paper and the reader is referred to \citet{kirk99}. As with the previous models a simplifying assumption of $K_e =1$ has been made.  The resulting plot is similar to Fig.~\ref{fig:fig4} and Fig.~\ref{fig:fig5}, and shows a significant increase in the scattering rate.

The electron distributions used are summarised in Table~\ref{tab:spectrum summary}.

\begin{table*}
 \begin{minipage}{160mm}
\caption{Summary of electron distributions considered.}
\label{tab:spectrum summary}
\begin{center}
% use packages: array
\begin{tabular}{cllccc} \hline
Figure & Distribution & Origin & $p$ & $\gamma_{min}$ & $\gamma_{max}$ \\  \hline
2 & Mono-energetic & pre-shocked & - & $10^6$ & $10^6$ \\ 
3 & Broad distribution & post-shock & 2.2 & $10^4$ & $10^7$ \\ 
4 & Narrow distribution & post-shock & 2.2 & $10^6$ & $10^7$ \\ 
5 & Adiabatic \citep{kirk99} & post-shock & 2.4 & $5.4\times10^5$ & $5.4\times10^7$ \\ 
6 & Radiative \citep{kirk99} & post-shock & 1.4 & $4.3\times10^5$ & $4.3\times10^7$ \\ \hline
\end{tabular}
\end{center}
\end{minipage}
\end{table*}

\section{Discussion}
\label{sec:discussion}

\subsection{The Influence of the IR excess}

\begin{figure}
\includegraphics[scale=.45]{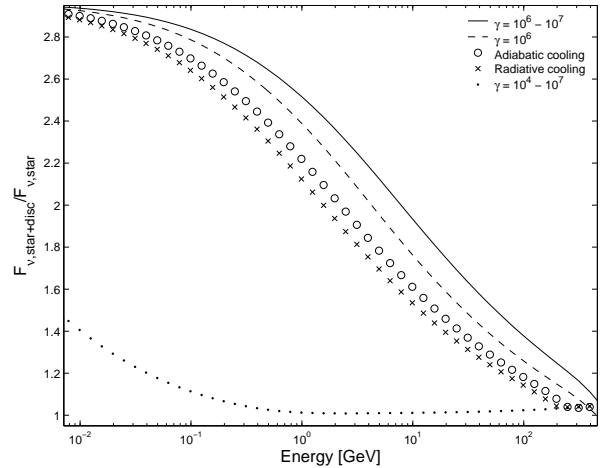}
 
 \caption{Fractional increase of the flux due to the inclusion for the IR excess for different electron distributions. The solid and dotted lines follow a power law distribution with $p=2.2$, the dashed line is for a mono-energetic pulsar wind, and the circle and crosses correspond to the electron distributions from \citet{kirk99} discussed in the text.  A disc radius of $50~R_*$ is assumed for all cases.}
 \label{fig:fig7}
\end{figure}

The results above show that for the case when higher energy electrons dominate the IC scattering, the gamma-ray spectrum can be influenced by the infrared excess.  Fig.~\ref{fig:fig7} shows the increase in the gamma-ray flux due to the inclusion of the IR excess as the ratio ${F_{\nu,star+disc}}/{F_{\nu,star}}$ versus energy for the different electron distributions considered.

The extent to which the gamma-ray spectrum is affected by the inclusion of the IR excess is dependent on which components of the Be star/disc spectrum are scattered in the Thomson limit.  
The peak in the stellar spectrum occurs at $\nu \sim (2 - 3) \times 10^{15}$~Hz (Fig.~\ref{fig:fig1}), and IC scattering will occur in the Thomson limit for electrons with Lorentz factors $\gamma \ll 10^5 \, (\nu/10^{15}\rmn{Hz})^{-1}$ (see equation~(\ref{eqn:thomlimit})). 
For the case of the broad electron energy distribution ($\gamma = 10^4 - 10^7$) the gamma-ray spectrum is dominated by electrons with Lorentz factors $\gamma = 10^4$ scattering photons from the peak of the stellar spectrum.   This scattering will occur in the Thomson limit and produce gamma-rays with energies
\[
\epsilon_1 \sim \gamma^2 h \nu \sim 10^9 \left(\frac{\gamma}{10^4}\right)^2 \left(\frac{\nu}{3\times10^{15}~\rmn{Hz}} \right)~\rmn{eV.}
\]
As a result, the major contribution at GeV energies is from the Thomson scattering of the whole stellar spectrum and the resulting gamma-ray spectrum is negligibly affected by the IR excess, as is shown in Fig.~\ref{fig:fig3} and Fig.~\ref{fig:fig7}.

For electrons with Lorentz factors $\gamma =10^5$, the scattering will occur in the Thomson limit only at frequencies $\nu \ll 10^{15}$~Hz.  As a result, scattering of photons from the peak in the stellar spectrum will produce gamma-rays in the Klein-Nishina limit, while photons from the IR excess will scatter in the Thomson limit and increase the gamma-ray flux at GeV energies.

In Fig.~\ref{fig:fig7} it is shown that, with the exception of the broad ($\gamma = 10^6- 10^7$) electron distribution, the gamma-ray flux increases by a factor $\ga 2$ at energies less than a few GeV.  The results show that the inclusion of the IR excess in the target photon distribution can influence the production of gamma-rays in \psrb, particularly at GeV energies.

\subsection{Variability of the Gamma-ray emission}

Since it is known that the circumstellar discs around Be stars are variable over periods of hundreds to thousands of days, the resulting change in the IR flux should also 
create variability in the GeV gamma-ray emission from \psrb. 

\begin{figure}
\includegraphics[scale=.45]{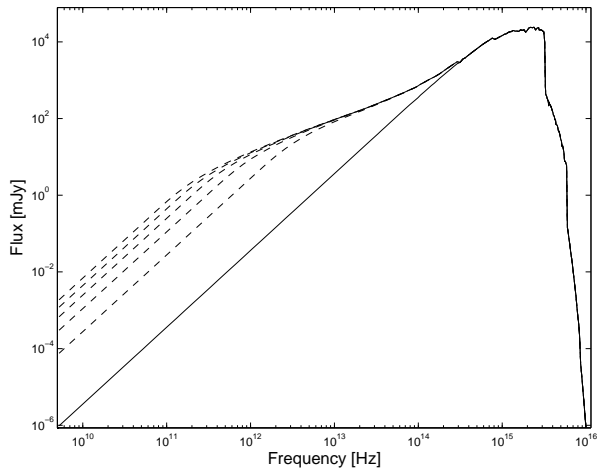}
 % broad_flux.eps: 0x0 pixel, 0dpi, 293.40x0.00 cm, bb=
 \caption{Broad energy distribution of SS~2883 predicted using the COG method. The solid line is the standard Kurucz model, while the dashed lines are the resulting IR excess for increasing disc sizes.  The disc sizes are (from the bottom to the top) 10, 20, 30, 40 and 50~$R_*$.  Only the disc radius is changed while the disc temperature and another parameters are the same as in Fig.~\ref{fig:fig1}.}
 \label{fig:fig8}
\end{figure}

\begin{figure}
  \includegraphics[scale=.45]{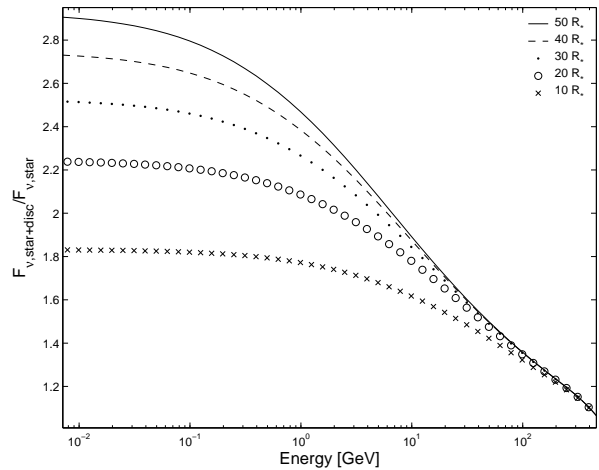}
% \includegraphics[scale=.45,bb=0   0   507   397]{gamma.eps}
 % gamma.eps: 2095434304x32637 pixel, -2147483648dpi, 0.00x0.00 cm, bb=
 \caption{Predicted increase in the IC flux for different disc radii, shown as the ratio of the modified flux to the initial flux. The disc sizes are 10, 20, 30, 40 and 50~$R_*$ from the bottom to the top, and the photon number distribution is calculated as is shown in Fig.~\ref{fig:fig8}. }
 \label{fig:fig9}
\end{figure}

A simple prediction can be made to show how the changing IR excess will modify the gamma-ray emission for \psrb.  By changing the disc radius parameter in the COG method (while keeping the other parameters constant i.e.\ $n$, $X_*$, $T_\rmn{disc}$ and $\theta$) the magnitude of the IR excess can be simulated for different size circumstellar discs in \psrb.   The resulting IR variability is shown in Fig.~\ref{fig:fig8}, where the lower solid line is a standard Kurucz model, and the dashed lines show the IR excess for increasing disc sizes.
The consequential increase in the IC flux is shown in Fig.~\ref{fig:fig9} as the ratio ${F_{\nu,star+disc}}/{F_{\nu,star}}$ versus energy.  The same disc sizes are used as in Fig.~\ref{fig:fig8} and the electron distribution follows the power law used in Fig.~\ref{fig:fig4}.  Fig.~\ref{fig:fig9} shows that by including the IR excess in the target photon distribution, the gamma-ray flux can increase by a factor of $\ga 2$
and show variability in the GeV energy range.

The modelling above assumes that the circumstellar disc only influences the photon distribution at IR wavelengths. However, it is known that the circumstellar disc also increases the magnitude of Be stars at optical wavelengths. This will greatly increase the total number of target photons and may result in an even greater possible variability of the gamma-ray flux during consecutive periastron passages. An increase in the Be star's flux at optical wavelengths could also lead to variability in the case of the broad electron distribution ($\gamma = 10^4 - 10^7$) as an increase in the flux near the peak of the stellar spectrum will lead to an increase in the total gamma-ray flux.

\subsection{Detectability of the gamma-ray modulation}

The increase in the gamma-ray flux due to the inclusion of the IR excess in an isotropic approximation will not be observable with current gamma-ray telescopes.  If the modelled fluxes are compared to the HESS spectrum the modified flux will fall below the \textit{Fermi} detection threshold.  A larger broad electron spectrum with more electrons around $\gamma \sim 10^4$ may produce a gamma-ray spectrum large enough to be be detected by \textit{Fermi}, but the variability due to the IR excess reported here would not be observable.

The analysis discussed here does however assume only isotropic scattering and the geometric size of the disc has not been considered.  Anisotropic modelling of the system may show a more pronounced effect, especially close to periastron where the circumstellar disc not only provides an IR excess but scattering will occur over the larger solid angle of the disc.

\section{Conclusion}
\label{sec:conclusion}

\psrb\ is a particularly important gamma-ray binary system since it contains an independently confirmed 48~ms pulsar compact object (as opposed to LS~5039 and LS~I~+61$^\circ$303 where the nature of the compact object must be inferred) and the gamma-ray emission is most likely created through the IC scattering of target photons from the optical companion via the electrons/positrons emanating from the pulsar wind. 
Previous modelling of the IC gamma-ray emission has not taken into account the IR excess created by the circumstellar disc and how these additional target photons will influence the IC scattering.

In this paper the \citet{lamers84} COG method was fitted to SS~2883 using optical, near-IR and mid-IR data available in literature and archives, under the assumption that the circumstellar disc has a radius of $R_\rmn{disc} = 50 R_*$, and that the Be star is a main sequence star with the parameters $T_{*} = 25000$~K,  $\log~g = 3.5$,  $M_* \simeq 10$~M$_{\odot}$ and $R_* \simeq 6$~R$_\odot$ based on the least squares fit performed in this paper and \citet{johnston94}.  The fit was used to calculate the target photon number density which was in turn used to calculate the IC spectrum for \psrb.  For a first approximation only isotropic scattering was considered in order to demonstrate the relevance of the IR excess.  

The modelling shows that the IR excess has a non-trivial influence on the IC spectrum, particularly at GeV energy gamma-rays where the flux increases by a factor $\ga 2$
for reasonable electron distributions. It has also been shown (since the circumstellar disc is variable) how changes in the IR excess will manifest as variability at gamma-ray energies. Consequently, observations of consecutive periastron passages may show differences in the GeV gamma-ray light curve as a result of a possibly changing IR excess.  

The results presented above have implications for other Be-XPBs and gamma-ray binary systems.  Previous modelling of the IC spectrum only considered a blackbody or mono-energy distribution for the target photons, however, the results above show that the IR excess from the circumstellar disc will have a non-trivial influence on the IC gamma-ray spectrum that must be accounted for when modelling \psrb.

\section*{Acknowledgements}

The authors are very grateful to O.\ de Jager for suggesting this study. This publication makes use of data products from the Two Micron All Sky Survey, which is a joint project of the University of Massachusetts and the Infrared Processing and Analysis Center/California Institute of Technology, funded by the National Aeronautics and Space Administration and the National Science Foundation.
This research made use of data products from the Midcourse Space Experiment. Processing of the data was funded by the Ballistic Missile Defense Organization with additional support from NASA Office of Space Science. This research has also made use of the NASA/ IPAC Infrared Science Archive, which is operated by the Jet Propulsion Laboratory, California Institute of Technology, under contract with the National Aeronautics and Space Administration.
BvS is funded by the South African Square Kilometre Array Project.
The authors would like to thank the anonymous reviewer for their comments which have improved this study.

\end{document}